\documentclass[sn-mathphys]{sn-jnl}% Math and Physical Sciences Reference Style
\jyear{2023}%

\theoremstyle{thmstyleone}%
\usepackage{subfig}
\theoremstyle{thmstyletwo}%
\theoremstyle{thmstylethree}%

\begin{document}
%%
%% The "title" command has an optional parameter,
%% allowing the author to define a "short title" to be used in page headers.
\title{Performance study of Partitioned Caches in Asymmetric Multi-Core processors}

%%
%% The "author" command and its associated commands are used to define
%% the authors and their affiliations.
%% Of note is the shared affiliation of the first two authors, and the
%% "authornote" and "authornotemark" commands
%% used to denote shared contribution to the research.
% \author{Murali Dadi}
% % \authornote{Both authors contributed equally to this research.}
% \email{muralidadi963@gmail.com}
% \affiliation{%
%   \institution{Indian Institute of Technology Madras}
% %   \streetaddress{Chennai}
%   \city{Chennai}
% %   \state{}
%   \country{India}
% %   \postcode{43017-6221}
% }
% % \orcid{1234-5678-9012}
% \author{Shubhang Pandey}
% % \authornotemark[1]
% \email{ee19s057@smail.iitm.ac.in}
% \affiliation{%
%   \institution{Indian Institute of Technology Madras}
% %   \streetaddress{Chennai}
%   \city{Chennai}
% %   \state{}
%   \country{India}
% %   \postcode{43017-6221}
% }

% \author{Aparna Behara}
% \affiliation{%
%   \institution{Indian Institute of Technology Madras}
% %   \streetaddress{Chennai}
%   \city{Chennai}
% %   \state{}
%   \country{India}
% %   \postcode{43017-6221}
% }
% \email{ee16d038@smail.iitm.ac.in}

% \author{T.G. Venkatesh}
% \affiliation{%
%   \institution{Indian Institute of Technology Madras}
% %   \streetaddress{Chennai}
%   \city{Chennai}
% %   \state{}
%   \country{India}
% %   \postcode{43017-6221}
% }
% \email{tgvenky@ee.iitm.ac.in}

\author{\fnm{Murali} \sur{Dadi}}\email{muralidadi963@gmail.com}

\author*{\fnm{Shubhang} \sur{Pandey}}\email{ee19s057@smail.iitm.ac.in}

\author{\fnm{Aparna} \sur{Behera}}\email{ee16d038@smail.iitm.ac.in}

\author*{\fnm{T G} \sur{Venkatesh}}\email{tgvenky@ee.iitm.ac.in}
% \equalcont{These authors contributed equally to this work.}

% \author[1,2]{\fnm{Third} \sur{Author}}\email{iiiauthor@gmail.com}
% \equalcont{These authors contributed equally to this work.}

\affil*{\orgdiv{Electrical Engineering Department}, \orgname{Indian Institute of Technology Madras}, \orgaddress{ \city{Chennai}, \country{India}}}

%%
%% By default, the full list of authors will be used in the page
%% headers. Often, this list is too long, and will overlap
%% other information printed in the page headers. This command allows
%% the author to define a more concise list
%% of authors' names for this purpose.
% \renewcommand{\shortauthors}{Murali Dadi et al.}

%%
%% The abstract is a short summary of the work to be presented in the
%% article.
\abstract

The current workloads and applications are highly diversified, facing critical challenges such as the Power Wall and the Memory Wall Problem. Different strategies over the multiple levels of Caches have evolved to mitigate these problems. Also, to work with such diversified applications, the Asymmetric Multi-Core Processor (AMP) presents itself as a viable solution. In this paper, we study the performance of L2 and Last Level Cache for different cache partitions against various AMP configurations. In addition, this study investigates the optimal cache partitioning for a collection of Multi-threaded benchmarks from PARSEC and SPLASH2 benchmark suites under medium-sized inputs. We have studied the effect of block replacement strategies and their impact on the key metrics such as total on-chip power consumption and L2 \& LLC Miss rates. Our study presents an intermediate cache design for AMPs between the two extremities of fully shared and fully private L2 \& LLC level Cache, which helps achieve the desired power values and optimal cache miss penalties.  

% \end{abstract}

%%
%% The code below is generated by the tool at http://dl.acm.org/ccs.cfm.
%% Please copy and paste the code instead of the example below.
%%
% \begin{CCSXML}
% <ccs2012>
%  <concept>
%   <concept_id>10010520.10010553.10010562</concept_id>
%   <concept_desc>Computer systems organization~Embedded systems</concept_desc>
%   <concept_significance>500</concept_significance>
%  </concept>
%  <concept>
%   <concept_id>10010520.10010575.10010755</concept_id>
%   <concept_desc>Computer systems organization~Redundancy</concept_desc>
%   <concept_significance>300</concept_significance>
%  </concept>
%  <concept>
%   <concept_id>10010520.10010553.10010554</concept_id>
%   <concept_desc>Computer systems organization~Robotics</concept_desc>
%   <concept_significance>100</concept_significance>
%  </concept>
%  <concept>
%   <concept_id>10003033.10003083.10003095</concept_id>
%   <concept_desc>Networks~Network reliability</concept_desc>
%   <concept_significance>100</concept_significance>
%  </concept>
% </ccs2012>
% \end{CCSXML}

% \ccsdesc[500]{Computer systems organization~Embedded systems}
% \ccsdesc[300]{Computer systems organization~Redundancy}
% \ccsdesc{Computer systems organization~Robotics}
% \ccsdesc[100]{Networks~Network reliability}

%%
%% Keywords. The author(s) should pick words that accurately describe
%% the work being presented. Separate the keywords with commas.

%%
%% This command processes the author and affiliation and title
%% information and builds the first part of the formatted document.
\maketitle
\section{Introduction}
\label{sec:introduction}
The current workloads and applications are highly diversified, facing critical challenges such as the Power Wall and the Memory Wall Problem. Different strategies over the multiple levels of Caches have evolved to mitigate these problems. Also, to work with such diversified applications, the Asymmetric Multi-Core Processor (AMP) presents itself as a viable solution. In this paper, we study the performance of L2 and Last Level Cache for different cache partitions of various AMP configurations. In addition, this study investigates the optimal cache partitioning for a collection of Multi-threaded benchmarks from PARSEC and SPLASH2 benchmark suites under medium-sized inputs. We have studied the effect of block replacement strategies and their impact on the key metrics such as total on-chip power consumption and L2 \& LLC Miss rates. Our study presents an intermediate cache design for AMPs between the two extremities of fully shared and fully private L2 \& LLC level Cache, which helps achieve the desired power values and optimal cache miss penalties.  

% So many power-related simulation tools CACTI, McPAT (Multi-Core power, area, timing), are also introduced to analyze and optimize the power. McPAT is an integrated power, area, and Timing tool used to model dynamic power. 

 There has been immense progress in recent times in designing high-performance and energy-efficient asymmetric Multi-Core processors. However, the trade-off between performance and power plays a crucial role in the processor design. With the aggressive scaling in IC Technology, power density $(W/cm^{2})$  has increased due to an increase in the number of transistors per unit area \cite{martin2014multicore}. 
%Even with high-end computers, power consumption, is a key consideration.  
In addition, the objective function (better performance or lower power consumption) may also change depending on the requirements and the operating conditions of a device. For example, in mobiles, energy needs to be optimized during idle periods for getting better battery life, while the performance needs to be prioritized during active times. LLCs are one of the processor's resources that significantly impact system performance and energy usage. So it becomes a significant challenge to handle LLCs efficiently \cite{martin2014multicore},\cite{Ref52}. Motivated by this point we set the aim of our paper to carry out  an extensive performance evaluation  of the LLC of AMPs as detailed below. Note that in the context of this paper, we have three cache levels and the terms L3 and LLC have been used interchangeably.

In this paper, we have 
studied the performance aspects of LLC in Asymmetric Multi-Core architectures. The high level overview of the paper is as follows.
% In Single-Core architecture, we have studied the effect of the size of the LLC and its associativity on the system performance. Primarily, the impact of the different replacement policies on the performance of the LLC under different configurations defined by its size and associativity has been examined.
In Multi-Core architecture, managing the shared LLCs is a critical task. Thus we have explored the effect of different configurations for L2 and LLCs which affects the critical system metrics such as L2 miss rate, L3 miss rate, and total power consumption of a Multi-Core architecture. Further in our study, we have investigated the impact on the power consumption of an Asymmetric Multi-Core processor due to different cores  (with different operating frequencies) and also different order of execution  (in-order or out-of-order). This forms the high level overview of our paper.

The remainder of this paper is organized as follows. Section \ref{section:lit} gives us a brief literature survey on the performance evaluation of the LLC. Then, section \ref{sec:study} presents the study related to the LLC and its corresponding simulation results. Inferences drawn from these simulation results and the concluding remarks are presented in Section \ref{sec:inf}. %Finally, the conclusion and the future scope related to our study are given in Section \ref{sec:con}.

\section{Literature Survey}
\label{section:lit}
This section reviews the existing literature works related to the study of performance and energy efficiency of the LLC.
Cache memories are often employed in microprocessors to increase the system performance, and thus these caches have been the subject of numerous studies\cite{Ref51, Ref52, Ref53,Ref54, montaner2012new}. The replacement policies of LLC   significantly affect the off-chip miss traffic and power consumption.
 Peneau et al. have studied how different LLC replacement strategies affect the system performance, and energy consumption \cite{Ref2}.
The asymmetric Multi-Core architectures are in high demand, and the existing replacement policies have significant challenges when implemented in Asymmetric Multi-Core systems. Ramtake et al. have studied the effect of Associativity on L1, L2 caches in a Multi-Core system concerning the cache hit ratio and IPC (Instructions per Cycle) \cite{Ref3}. The modern VLSI chips integrate larger caches onto the processor, and managing such larger cache sizes has considerable overhead.
Wu et al. have proposed a machine learning based   management scheme for the shared LLC  in chip multiprocessors \cite{qu}.
Jang et al. \cite{Ref4} have suggested a cache design for larger LLCs, so that good performance is attained even with high granularity.  Anandkumar et al. have proposed a new hybrid cache replacement strategy for heterogeneous Multi-Cores that combines LRU, and  LFU replacement policies \cite{Ref5}. Heterogeneity in a Multi-Core system can be achieved by changing individual core frequencies, cache sizes, and other cache parameters. Silva et al. have investigated the benefits of having various cache sizes in HMPs (Heterogeneous Multi-Core Processors) and how a scheduling technique can explore such benefits to reduce the overall miss rate \cite{Ref6}.
The LLC   in a modern chip-multiprocessor  (CMP) is typically shared by all the cores. Processors use the shared caches more frequently; therefore, eviction of the shared data causes more cache misses\cite{j1}. Thus, to efficiently utilize the shared LLC on a CMP, Sato et al. have proposed cache partitioning to protect the shared data by reducing unnecessary evictions \cite{Ref7}. This approach separates shared and private data and uses cache partitioning to give each type of data its own cache space. Several research works have been done on the partitioning of the shared LLC to improve system performance. But they all miss the heterogeneity in the spatial locality of different applications. Gupta et al. \cite{Ref8} showed how leveraging spatial locality allows significantly more effective cache sharing. They have highlighted that when large block size is used, the cache capacity requirements of many memory-intensive applications can be dramatically lowered, allowing them to give more capacity to other workloads effectively.  In CMPs (Chip Multi Processors), private LLC provides a better access latency than shared caches. But more private caches result in replication of shared data, leading to under-utilization of total net capacity of cache, thus decreasing the overall hit ratio.
A part of the work performed by Chen et al. on private and shared caches highlight the above mentioned issues both in terms of performance and energy \cite{chen2018energy}. 
To handle the problem mentioned above, Yuan et al. have proposed a new cache management technique that improves the performance of a CMP using the private LLC \cite{Ref9}. Sibai et al. have discussed issues related to sharing and privatizing second and third-level cache memories in homogeneous Multi-Core architectures \cite{sibai}.

%When we integrate hundreds of cores on a single chip, it becomes very complex and expensive to have an efficient cache coherence protocol that maintains the consistency of shared data. Kaur et al. have presented a cache coherency management technique for the non-coherent cache architectures that uses an automatic parallelizing compiler \cite{Ref10}.  The traditional Snoopy-based coherence protocols are not scalable, and they need much higher bandwidth. The performance of the shared memory multiprocessors is further increased by using an efficient method for addressing the cache coherency.  Asaduzzaman et al. have used the snoopy and directory techniques for designing a hybrid cache coherence protocol which achieves an improved sharer group mechanism \cite{Ref11}. 

 Although caches can significantly boost system performance, they use a considerable amount of overall system power. Chakraborty et al. \cite{Ref12} have analyzed the effect of LLC on the chip temperature, and they have proposed a new policy that resizes on-chip LLC at run time so that the leakage power consumption is reduced. 
%  When establishing cache coherence, the most important factors to consider are performance, energy, and scalability. Joshi et al. have investigated the energy requirement and performance of different snoopy-based, and directory-based cache coherence protocols \cite{Ref14}.
%  Saez et al. have  presented an Asymmetry aware completely fair scheduling (ACFS) scheme to address the fairness among all the applications in single-ISA AMPs and also provide overall throughput improvement \cite{saez2017towards}.
%  On similar lines for fairness scheduling for various applications in single-ISA AMPs, \cite{contentionawarefairness} proposes CAMPS (Contention Aware Scheduling for AMPs), which does not require any external hardware or prediction models, for addressing the contention issues usually observed on shared resources like the LLC and the memory bus, the completely OS-level scheduler provides 11\% better fairness and even better throughput. 
%Sustran et al. present the idea for migrating transactions from one type of cores to the other type based, these cores they refer to as "big" and "small", as they differ to each other on some microarchitectural aspects. Theiir algorithm for migrating transactions from small cores to big core improved the overall execution performance\cite{sustran9422791}. 

From the above literature works, we observed that the performance of the LLC concerning the Multi-Core processors was investigated in-depth.
However the combined effect of sharing/partitioning both the L2 and LLC along with the effect of replacement needs further study.
Finally, the performance of the LLC in the context of asymmetric multi-core processors has not received much attention. To fill this gap, we have done an extensive performance study of the LLC, primarily concentrating on the heterogeneity of the processors.  
\\

% The major contributions of our paper are as follows.
% \begin{enumerate}
% \item The size and associativity of the LLC are varied, and its effect on the system performance is studied.
% \item We have studied the behavior of the key metrics of Asymmetric Multi-Core architecture, such as L2 miss rate, L3 miss rate, and total on-chip power consumption. 
% \end{enumerate}

The unique features of our paper are as follows.
\begin{enumerate}
\item While \cite{sibai} evaluates the performance of shared/private LLC for homogeneous Multi-Core processors, we concentrate on the same problem but for the case of  heterogeneous Multi-Core architectures.
\item In most of the papers such as \cite{Ref6}, the heterogeneity of the Multi-Core processors is introduced either by varying cache size or by varying block size. A unique feature of our paper is that we have introduced the heterogeneity by varying the core frequency as well as by varying the order of execution (either in-order or out-of-order execution).
\item Another novel feature of our work is that we
have extensively studied the performance of different  configurations of AMPs that differ by the way the L2 and LLC are partitioned and arrive at the optimal configuration.
\end{enumerate}

\section{Performance trade-off in Asymmetric Multi-Core Architectures (AMPs)} \label{sec:study}
\begin{table*}[!ht]
\footnotesize
\centering
\caption{L2 and L3 Cache details for different Configurations}
\label{tab:9onfigs}
\begin{tabular}{|p{2cm}|p{3cm}|p{3cm}|p{3cm}|}
 \hline
\textbf{Configuration Index} & \textbf{$L2$ Cache Details} & \textbf{$L3$ Cache Details} & \textbf{Core Frequencies} \\
 \hline 
 \hline
 Configuration $1$ & All $16$ cores sharing $2048KB$ of $L2$ cache & All $16$ cores sharing $8192KB$ of $L3$ Cache  & All $16$ cores are Out-of-order, running at $2.66GHz$ \\ 
 \hline
  Configuration $2$ & Two sets of $8$ cores and each set sharing $1024KB$ of $L2$ cache &  all cores sharing $8192KB$ of $L3$ Cache  &   All cores are Out-of-order, running at $2.66$GHz\\
  \hline
  Configuration $3$ & Four sets of $4$ cores, and each set sharing $512KB$ of $L2$ cache &   all cores sharing $8192KB$ of $L3$ Cache  &   All cores are Out-of-order, running at $2.66$GHz\\
\hline
  Configuration $4$ &  Eight sets of $2$ cores, and each set sharing $256KB$ of $L2$ cache &   all cores sharing $8192KB$ of $L3$ Cache  &   All cores are Out-of-order, running at $2.66GHz$\\
\hline
  Configuration $5$ &   Each core with private $L2$ cache of $128KB$ &    all cores sharing $8192KB$ of $L3$ Cache  &   All cores are Out-of-order, running at $2.66$GHz\\
\hline
  Configuration $6$ &  Each core with private $L2$ Cache of size $128KB$ &  Two sets of $8$ cores and each set sharing cache of size $4096KB$  &   Core $0-7$, in-order, speed $1GHz$. Cores $8-15$ out-of-order, speed $2.66GHz$ \\
\hline
  Configuration $7$ &  Each core with private $L2$ Cache of size $128KB$ & Four sets of $4$ cores, and each set sharing Cache of size $2048KB$  &   Cores $0-7$, in-order, speed  $1GHz$. Cores $8-15$   out-of-order, speed $2.66GHz$ \\
\hline
  Configuration $8$ &  Each core with private $L2$ Cache of size $128KB$ &    Eight sets of $2$ cores, and each set sharing Cache of size $1024KB$  &                   Cores $0-7$ in-order, speed $1GHz$. Cores $8-15$ out-of-order, speed $2.66GHz$ \\
\hline
  Configuration $9$ &  Each core with private $L2$ Cache of size $128KB$ &   each core with private  $L3$ Cache of size $512KB$  &  Cores $0-7$ in-order, speed $1GHz$ ,
                 Cores $8-15$ out-of-order, speed $2.66GHz$ \\
\hline
\end{tabular}

\end{table*}
In this section, we study the performance trade-off in Asymmetric Multi-Core architecture. Considering both performance and power as primary concerns, we have implemented nine different configurations. In configurations $1$ to $5$, all cores are powerful cores with Out-of-order execution and an operating frequency of 2.66$GHz$. But in the configurations $6$ to $9$, we have introduced Asymmetricity by changing the order of execution of cores (In-order or Out-of-order) and the frequency of operation of each core (1$GHz$ or  2.66$GHz$). In these configurations, cores $0-7$ are In-order cores running at 1$GHz$, whereas $8-15$ are Out-of-order cores running at 2.6$6GHz$. When Out-of-order and In-order cores share a cache memory, there is a chance that Out-of-order cores may occupy a large portion of the cache, which further degrades the performance of In-order cores running at a lower frequency. So we gradually partitioned the LLC among the cores from fully shared to fully private. The remaining details of all the nine configurations are given in Table \ref{tab:9onfigs}. The architecture used for our study, along with the corresponding simulation setup, is as given below.
\subsection{Architecture Studied}
\begin{figure}[h]
\vspace{1cm}
    \centering
    \includegraphics[width=0.75\textwidth]{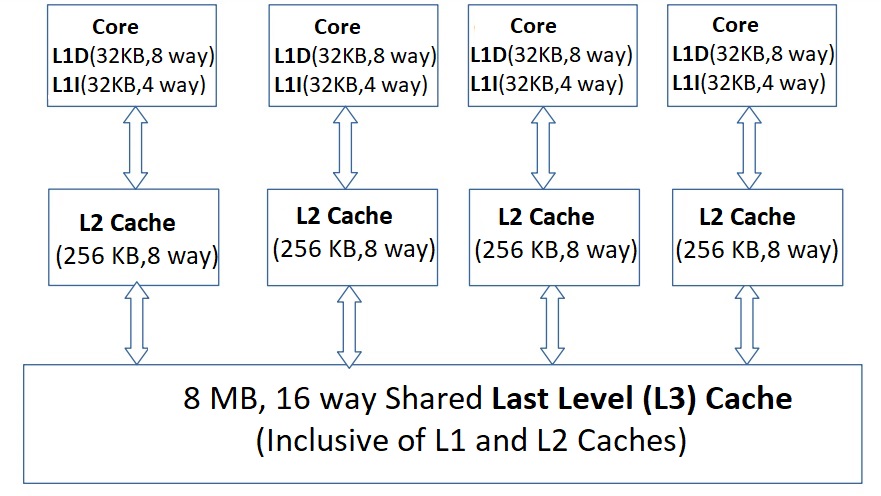}
    \caption{Cache hierarchy of Nehalem architecture with $4$ cores }
    \label{fig:Nehalem}
\end{figure}

%\textcolor{red}{better name the study}
We have referred to the Nehalem architecture \cite{singhal2008inside}, one of the most successful processor architectures introduced by \textsl{Intel}.  
%\textsl{Nehalem} is a dynamically scalable and design-scalable microarchitecture. It dynamically manages cores, threads, cache, and power to deliver outstanding energy efficiency and performance on demand.
 %\subsubsection*{An overview of Nehalem Architecture}
Nehalem has scalable performance for from $1$ to $16$ (or more) threads and from $1$ to $8$ (or more) cores. It contains scalable and configurable system interconnects and also contains an integrated memory controller. The three-level cache hierarchy of this microarchitecture which is shown in Fig \ref{fig:Nehalem} consists of $64$KB of $L1$ cache with $32$KB of data cache and $32$KB of the instruction cache. Further, it has $256$KB of $L2$ cache per core (private cache) for handling data and instructions. Finally, it has a fully inclusive and fully shared LLC of size $8$MB where all applications can use the entire cache. Nehalem has a more out-of-order window and scheduler size, which helps it identify more independent operations that can run in parallel. It also has larger-sized buffers in the core to ensure that they do not limit the performance.  

\subsection{Simulator and Workloads used}
We have used $Sniper $ v7.3  simulator for our study \cite{carlson2014aeohmcm}. It is an accurate high-speed $X86$ based simulator suitable for exploring different heterogeneous Multi-Core architectures. This simulator also provides us high-speed timing simulations for the multi-threaded, multi-program workloads and shared-memory applications.
 %In addition to these many features, it has the best base simulation infrastructure to simulate a more extensive set of workloads on more recent simulated hardware. 
 We have used five different workloads in this study from the PARSEC Benchmarks Suite \cite{parsec} and SPLASH2 Benchmark Suite \cite{splash2}. They are PARSEC-Bodytrack, PARSEC-freqmine,  SPLASH2-barnes, PARSEC-fluidanimate, and SPLASH2-radiosity.
         Further  \textsl{McPAT} (Multicore Power, Area, and Timing) framework is integrated with Sniper for modeling the power and area aspects of many-core architectures. We have used McPAT v1.0 \cite{mcpat} in this study to get the different power consumption values of the processor.\\
         
\subsection{Simulation Results} \label{sec:la}
\begin{figure}[t]
\centering     %%% not \center
\subfloat{\label{fig:l2_9configs}\includegraphics[width=0.5\textwidth]{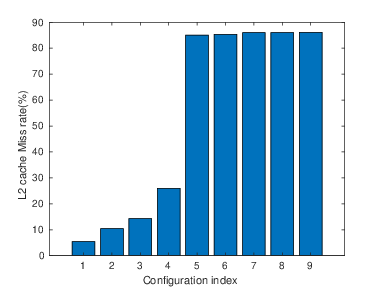}}\hfill
\subfloat{\label{fig:l3_9configs}\includegraphics[width=0.5\textwidth]{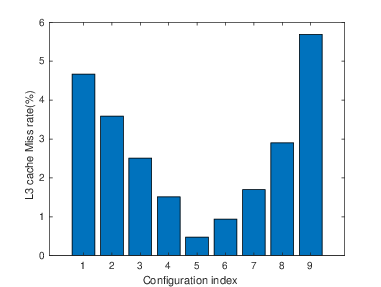}}\hfill
\subfloat{\label{fig:pow_9configs}\includegraphics[width=0.5\textwidth]{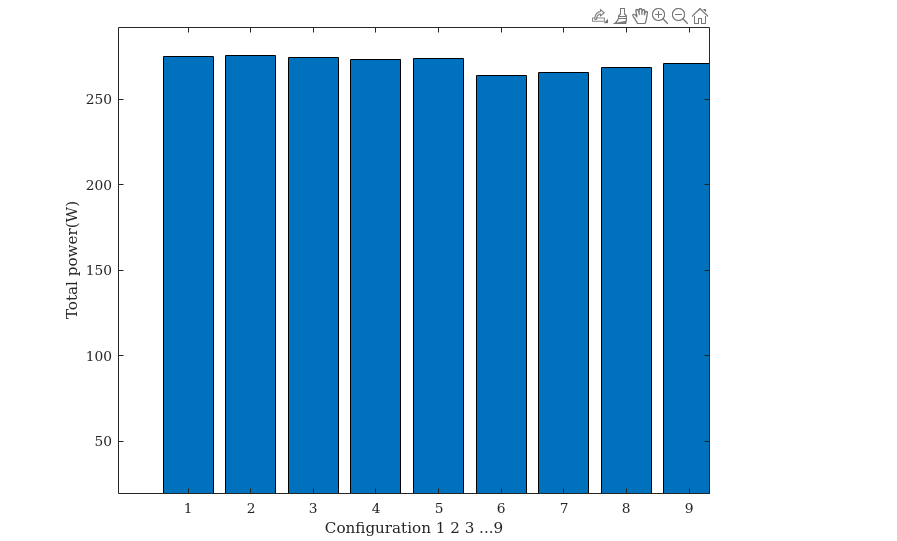}}
\caption{Performance trade-off in Heterogeneous Multi-Core Architecture}
\label{fig:hetro}
\end{figure}

We have simulated all nine different configurations in the initial stage as mentioned in Table \ref{tab:9onfigs}. We compared all these configurations using their L2 miss rate, L3 miss rate, and total power consumption as the metrics for comparison. The consolidated results are given in Tables \ref{tab:l2_9configs},\ref{tab:l3_pow_9configs} and presented separately in the sub-figures of Fig \ref{fig:hetro} (a),  (b) and  (c) for better understanding. Except for $L2$ and $L3$ Cache levels,  all the remaining parameters are common in all the configurations, which are as mentioned in Table \ref{tab:nehalem}
\begin{table}[!h]
\footnotesize
\centering
\caption{Configuration details}
\label{tab:nehalem}
\begin{tabular}{|p{4cm}|p{4cm}|}
% {|c|c|c|}
 \hline
  Number of Cores & $16$ \\   
  Block Size & $64$ Bytes \\
  Cache Coherence Protocol & MESI\\
  Replacement Policy & LRU \\
  %Order Of Execution & InO (Core $0-7$, OoO  (core $8-15$)\\
  $L1$ size & $32KB (I)$, $32KB (D)$\\
  $L1$ shared cores & $1$ (private to each core)\\
  $L1$ Associativity & $8 (D)$, $4 (I)$ \\
  $L2$ Size & $2048KB$ (Total)\\
  $L2$ Associativity & $8$\\
  $L3$ Size & $8192KB$ (Total)\\
  $L3$ Associativity & $16$\\
\hline
\end{tabular}

\end{table}
From Fig. \ref{fig:l2_9configs}, we observe that from configuration 1 to configuration 5, the $L2$ cache is changing from fully shared among all cores to fully private to each core. When we have more private caches, the chance of coherence misses is more, which increases the miss rate. 
Hence $L2$ Miss rate is increasing when moving from fully shared to fully private caches. But $L2$ cache is private to each core from configuration 5 to configuration 9, where the miss rate is almost the same.
% \begin{table*}[!hb]
% \footnotesize
% \centering
% \begin{tabular}{|p{1.6cm}|p{3.2cm}|p{1.3cm}|p{3.5cm}|p{1.3cm}|p{1.4cm}|}
% % {|c|c|c|}
%  \hline
%  Configuration Index &  Total $L2$ Misses / Total $L2$ Accesses & $L2$  Miss rate  ($\%$) & Total $L3$ misses/Total $L3$ Accesses & $L3$ Miss rate $ (\%)$ & Total Power (W)  \\ 
%  \hline
%   Configuration $1$ & $1,775,452/32,658,514$ & $5.43$ & $83,872/1,792,716$ & $4.67$ &$275.238$\\   Configuration $2$ & $2,414,712/23,034,264$ & $10.48$ & $87,401/2,430,864$ & $3.59$ & $275.449$\\   Configuration $3$ & $3,464,822/24,173,825$ & $14.33$ & $87,561/3,481,529$ & $2.51$ & $274.293$\\   Configuration $4$ & $5,413,832/20,838,462$ & $25.98$ & $82,067/5,435,954$ & $1.51$ & $273.325$\\   Configuration $5$ & $17,851,930/20,985,015$ & $85.06$ & $85,948/17,871,357$ & $0.48$ & $273.718$\\   Configuration $6$ & $17,945,030/20,985,015$ & $85.36$ & $169,125/17,984,595$ & $0.94$ & $263.471$\\  
%  Configuration $7$ & $18,105,468/21,045,636$ & $85.99$ & $309,358/18,176,373$ & $1.7$ & $265.63$\\   Configuration $8$ & $18,096,468/21,045,663$ & $85.99$ & $527,606/18,178,885$ & $2.9$ & $268.418$\\   Configuration $9$ & $18,013,786 / 20,914,704$ & $86.13$ & $1,033,091/18,161,730$ &$5.688$ & $270.978$\\ 
% \hline
% \end{tabular}
% \caption{$L2$ cache Miss rate for all configurations}
% \label{tab:l2_l3_pow}
% \end{table*}
\begin{table}[!h]
\footnotesize
\centering
\caption{$L2$ cache Miss rate for all configurations}
\label{tab:l2_9configs}
\begin{tabular}{|p{2cm}|p{3.5cm}|p{1.3cm}|}
% {|c|c|c|}
 \hline
 Configuration Index &  Total $L2$ Misses / Total $L2$ Accesses  & $L2$ Miss rate  \\ 
 \hline
  Configuration $1$ & $1,775,452/32,658,514$ & $5.43$  \\   Configuration $2$ & $2,414,712/23,034,264$ & $10.48$ \\   Configuration $3$ & $3,464,822/24,173,825$ & $14.33$ \\   Configuration $4$ & $5,413,832/20,838,462$ & $25.98$ \\   Configuration $5$ & $17,851,930/20,985,015$ & $85.06$ \\   Configuration $6$ & $17,945,030/20,985,015$ & $85.36$ \\  
 Configuration $7$ & $18,105,468/21,045,636$ & $85.99$ \\   Configuration $8$ & $18,096,468/21,045,663$ & $85.99$ \\   Configuration $9$ & $18,013,786 / 20,914,704$ & $86.13$ \\ 
\hline
\end{tabular}

\end{table}

\begin{table}[!h]
\footnotesize
\centering
\caption{$L3$ cache Miss rate and Power for all configurations}
\label{tab:l3_pow_9configs}
\begin{tabular}{|p{2cm}|p{3.2cm}|p{1.3cm}|p{1.4cm}|}
% {|c|c|c|}
 \hline
 Configuration Index &  Total $L3$ misses/Total $L3$ Accesses & $L3$ Miss rate $ (\%)$ & Total Power (W)  \\ 
 \hline
  Configuration $1$  & $83,872/1,792,716$ & $4.67$ &$275.238$\\   Configuration $2$  & $87,401/2,430,864$ & $3.59$ & $275.449$\\   Configuration $3$  & $87,561/3,481,529$ & $2.51$ & $274.293$\\   Configuration $4$  & $82,067/5,435,954$ & $1.51$ & $273.325$\\   Configuration $5$  & $85,948/17,871,357$ & $0.48$ & $273.718$\\   Configuration $6$  & $169,125/17,984,595$ & $0.94$ & $263.471$\\  
 Configuration $7$  & $309,358/18,176,373$ & $1.7$ & $265.63$\\   Configuration $8$  & $527,606/18,178,885$ & $2.9$ & $268.418$\\   Configuration $9$ & $1,033,091/18,161,730$ &$5.688$ & $270.978$\\ 
\hline
\end{tabular}

\end{table}
% \newpage
% II.$L3$ cache Miss rate of all configurations\\
% \begin{figure}[h]
%   \subfloat{
%     \includegraphics[width=\linewidth]{l3_9configs_table.png}
%             }
%   \subfloat{
%   \includegraphics[width=\linewidth]{L3_9configs.png}
%             }
  
%     \label{fig:my_label}
% % \end{figure}   \\
% \textbf{OBSERVATIONS}\\
From Fig. \ref{fig:l3_9configs}, we can observe that when as we go from configuration $1$ to $5$, the L3 miss rate is decreasing even if the L3 cache is shared in all these cases. If we observe the number of total misses to the L3 cache in configurations $1$ to $5$ in the Table \ref{tab:l3_pow_9configs}, they are almost the same. Due to the increasing miss rate of L2, the number of accesses to the L3 cache increases, reducing the overall  L3 miss rate.
L3 cache is changing from a fully shared cache to a fully private cache in configurations $5$ to $9$, increasing the miss rate due to increased data inconsistency  (coherent misses). In addition, when the number of private caches is increased, there is a chance for replicating the same data, which also increases the total miss rate due to inefficient utilization of net cache capacity.
% \begin{figure}[h]
%   \subfloat{
%     \includegraphics[width=\linewidth]{power_9configs_table.png}
%             }
%   \subfloat{
%   \includegraphics[width=\linewidth]{POWER_9CONFIGS.png}
%             }
  
%     \label{fig:my_label}
% \end{figure}   \\
 From Fig. \ref{fig:pow_9configs}, we observe that in  Configurations $1$ to $5$, all cores are running at $2.66$GHz (Out-of-Order), so the total power consumed is almost the same in all the cases. But configurations 6 to 9 have eight cores running at 1GHz (In-order), and the other eight cores are running at 2.66GHz (Out-of-order), which gives low power consumption. So the power consumption is reduced when we go for asymmetric Multi-Cores. However, as we go from configurations 5 to 9, the power consumption increases due to the increase in L3 miss rate, resulting in power-intensive off-chip main memory access.

\subsubsection{Configurations $1,6,9$ with different replacement  policies and with different workloads}
\label{section:config169}
% 1. Block diagram of configuration $1$:
% \textcolor{green}{we can remove these configuration figures}

% \begin{figure}
% \centering     %%% not \center
% \subfloat[Block diagram of configuration $1$]{\label{fig:config1_block}\includegraphics[width=120mm]{config1.png}}
% \subfloat[Block diagram of configuration $5$]{\label{fig:config5_block}\includegraphics[width=120mm]{config5.png}}
% \subfloat[Block diagram of configuration $9$]{\label{fig:config9_block}\includegraphics[width=120mm]{config9.png}}
% \caption{Configuration $1,5,9$ with different replacement  policies and with different workloads}
% \label{fig:config}
% \end{figure}

% \begin{figure}[h]
%     \centering
%     \includegraphics[width=\linewidth]{config1.png}
%     \caption{Block diagram of configuration $1$}
%     \label{fig:config1_block}
% \end{figure}

% \begin{figure}[h]
%   \subfloat{
%     \includegraphics[width=\linewidth]{config5.png}  }
%   \subfloat{
%   \includegraphics[width=\linewidth]{config9.png}
  
%             }
  
%     \label{fig:my_label}
% \end{figure}

% \newpage

\begin{figure}[!h] 
\centering     %%% not \center
\subfloat{\label{fig:lru_l2}\includegraphics[width=0.5\textwidth]{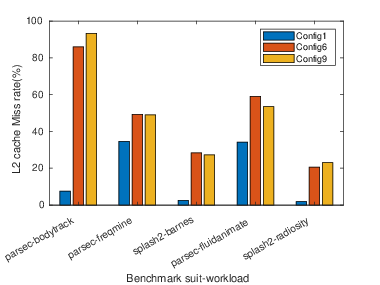}}\hfill
\subfloat{\label{fig:lru_l3}\includegraphics[width=0.5\textwidth]{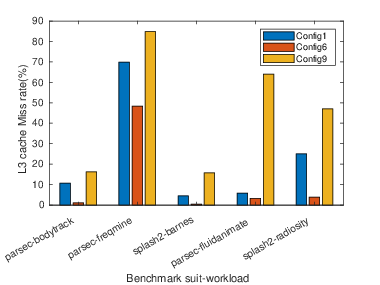}}\hfill
\subfloat{\label{fig:lru_pow}\includegraphics[width=0.5\textwidth]{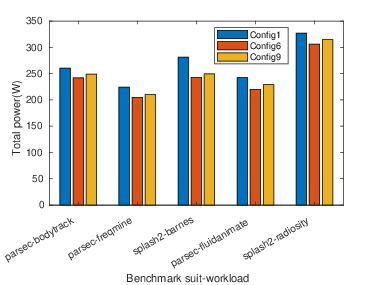}}
\caption{Configurations $1,6,9$ with LRU replacement policy}
\label{fig:lru_159}
\end{figure}

\begin{figure}[!h]
\centering     %%% not \center
\subfloat{\label{fig:mru_l2}\includegraphics[width=0.5\textwidth]{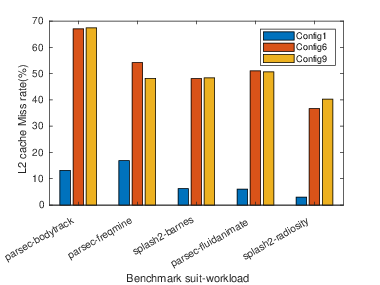}}\hfill
\subfloat{\label{fig:mru_l3}\includegraphics[width=0.5\textwidth]{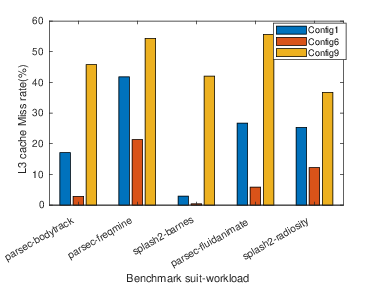}}\hfill
\subfloat{\label{fig:mru_pow}\includegraphics[width=0.5\textwidth]{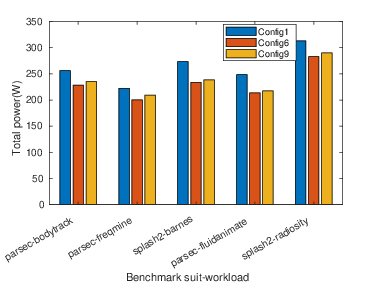}}
\caption{Configurations $1,6,9$ with MRU Replacement policy}
\label{fig:Mru_159}
\end{figure}

\begin{figure}[!h]
\centering     %%% not \center
\subfloat{\label{fig:ro_l2}\includegraphics[width=0.5\textwidth]{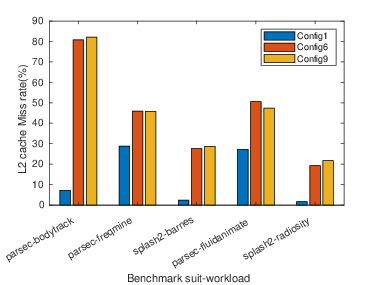}}\hfill
\subfloat{\label{fig:ro_l3}\includegraphics[width=0.5\textwidth]{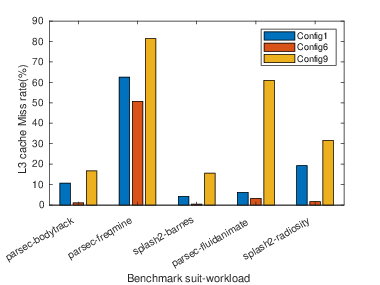}}\hfill
\subfloat{\label{fig:ro_pow}\includegraphics[width=0.5\textwidth]{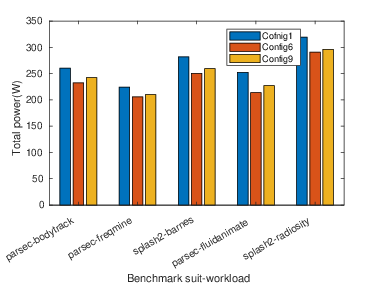}}
\caption{Configuration $1,6,9$ with Round Robin Replacement policies}
\label{fig:ROUND_159}
\end{figure}

% In this study, we have fixed the coherence protocol as MESI. The reason being the previous study has shown that the MESI protocol offers better performance and power efficiency. However,
In section \ref{sec:la}, we have encountered a trade-off between L3 miss rate and total power consumption of Asymmetric Multi-Core architecture. Using configuration $1$, configuration $6$, and configuration $9$, we want to investigate this trade-off further with three different replacement policies (LRU, MRU, Round Robin) using five different workloads (PARSEC-bodytrack, PARSEC-freqmine, SPLASH2-barnes, PARSEC-fluidanimate, SPLASH2-radiosity).
%\subsubsection*{LRU replacement policy}
We have examined the L2 miss rate, L3 miss rate, and total power consumption in configurations 1,6,9 with the five different workloads mentioned above. Corresponding results for the above mentioned simulation using three different replacement policies  are  shown in Fig. \ref{fig:lru_159}  ( LRU case),  Fig. \ref{fig:Mru_159}  (  MRU case) and Fig. \ref{fig:ROUND_159}  ( Round Robin case).
% Replacement policy is fixed with LRU in all the cases and run 1,5,9 configurations with different workloads. Fig. \ref{fig:lru_159} gives the corresponding re
% I.$L2$ Cache Miss rate \\

% \begin{figure}[h]
%   \subfloat{
%     \includegraphics[width=\linewidth]{l2_lru_table.png}
    
%             }
%             \\
%             \\
%   \subfloat{
%   \includegraphics[width=\linewidth]{l2_lru.png}

%             }
  
%     \label{fig:my_label}
% \end{figure}

% \newpage

% II.$L3$ Cache Miss rate\\
% \begin{figure}[h]
%   \subfloat{
%     \includegraphics[width=\linewidth]{l3_lru_table.png}
    
%             }
%             \\
%             \\
%   \subfloat{
%   \includegraphics[width=\linewidth]{l3_lru.png}

%             }
  
%     \label{fig:my_label}
% \end{figure}

% \newpage

% III.Total Power\\
% \begin{figure}[h]
%   \subfloat{
%     \includegraphics[width=\linewidth]{power_lru_table.png}
    
%             }
%             \\
%             \\
%   \subfloat{
%   \includegraphics[width=\linewidth]{power_lru.png}

%             }
  
%     \label{fig:my_label}
% \end{figure}

% \newpage
%\subsubsection*{{MRU} replacement policy}
%We have examined the L2 miss rate, L3 miss rate, and total power consumption in configurations 1,5,9 against the replacement policy MRU with the five different workloads mentioned earlier. Corresponding results are shown in Fig. \ref{fig:Mru_159}
% I.$L2$  Cache miss rate\\

% \begin{figure}[h]
%   \subfloat{
%     \includegraphics[width=\linewidth]{l2_mru_table.png}
    
%             }
%             \\
%             \\
%   \subfloat{
%   \includegraphics[width=\linewidth]{l2_mru.png}

%             }
  
%     \label{fig:my_label}
% \end{figure}

% \newpage

% II.$L3$  Cache miss rate\\

% \begin{figure}[h]
%   \subfloat{
%     \includegraphics[width=\linewidth]{l3_mru_table.png}
    
%             }
%             \\
%             \\
%   \subfloat{
%   \includegraphics[width=\linewidth]{l3_mru.png}

%             }
  
%     \label{fig:my_la/subel}
% \end{figure}

% \newpage

% II.Total power\\

% \begin{figure}[h]
%   \subfloat{
%     \includegraphics[width=\linewidth]{power_mru_table.png}
    
%             }
%             \\
%             \\
%   \subfloat{
%   \includegraphics[width=\linewidth]{power_mru.png}

%             }
  
%     \label{fig:my_label}
% \end{figure}

% \newpage
%\subsubsection*{Round Robin replacement policy}
%We have examined the L2 miss rate, L3 miss rate, and total power consumption in configurations 1,5,9 against the replacement policy Round Robin with the five different workloads mentioned above. Corresponding results are as shown in Fig. \ref{fig:ROUND_159}
% I.$L2$  Cache miss rate\\

% \begin{figure}[h]
%   \subfloat{
%     \includegraphics[width=\linewidth]{l2_roundrob_table.png}
    
%             }
%             \\
%             \\
%   \subfloat{
%   \includegraphics[width=\linewidth]{l2_roundrob.png}

%             }
  
%     \label{fig:my_label}
% \end{figure}

% \newpage

% II.$L3$  Cache miss rate\\
% \begin{figure}[h]
%   \subfloat{
%     \includegraphics[width=\linewidth]{l3_roundrob_table.png}
    
%             }
%             \\
%             \\
%   \subfloat{
%   \includegraphics[width=\linewidth]{l3_roundrob.png}

%             }
% %The current workloads and applications are highly diverse in nature. These applications face critical challenges such as the ILP Wall, Power Wall, Frequency Wall, and Memory Wall Problem.  The Asymmetric Multi-Core Processors  (AMP) presents itself as a viable solution to address these  diversified workloads and applications. We realize that the different Cache levels are greatly affected because of

%     \label{fig:my_label}
% \end{figure}

% \newpage
% III.Total Power\\
% \begin{figure}[h]
%   \subfloat{
%     \includegraphics[width=\linewidth]{power_roundrob_table.png}
    
%             }
%             \\
%             \\
%   \subfloat{
%   \includegraphics[width=\linewidth]{power_roundrob.png}

%             }
  
%     \label{fig:my_label}
% \end{figure}
Configuration 6 is better for both in terms of L3 miss rate and power. We observe that a trade-off exists between the L3 miss rate and total power consumption from the above figure. Independent of the replacement policy and the workloads used when we go from configuration $1$ to configuration $9$, total power consumption decreases significantly in all the cases, but the $L3$ miss rate increases. The LRU  and Round Robin replacement policies give a better L2 miss rate when compared to the MRU policy. But we need to consider the L3 misses over the L2 misses because the misses to the L3 cache go to the main memory, increasing the overall latency and energy consumption.
%From configuration 1 to configuration 9, we observed the 8\%, 16\%, 14\% increase in L3 miss rate and   7.4\%, 8.8\%, and 7\% reduction in total power consumption in LRU, MRU, and Round Robin replacement policies, respectively.

 For the LRU replacement policy, when the configuration is changed from $1$ to $6$, the L3 miss rate and power consumption are reduced by $11.6\%$ and  $9.058\%$, respectively. Further, when the configuration changes from $6$ to $9$, both the L3 miss rate and power consumption increase by $27.25\%$ and  $3.06\%$, respectively. When the configuration is changed from $1$ to $9$, the L3 miss rate increases by $13.25\%$, and power consumption reduces by $6.28\%$. In MRU replacement policy, from configuration $1$ to configuration $6$, L3 miss rate reduces by 14.2\%, and power consumption reduces by $11.752\%$. Later from configuration $6$ to configuration $9$, both the L3 miss rate and power consumption increases by $32\%$ and  $2.75\%$ respectively.  From configuration $1$ to configuration $9$, the L3 miss rate is increased by 20.25\%, and power consumption is reduced by 9.3\%.  In the Round Robin replacement policy, from configuration $1$ to configuration $6$, the $L3$ miss rate is reduced by $7.6\%$, and $10.81\%$ reduction is observed for power consumption. From configuration $6$ to configuration $9$, the L3 miss rate increases by 28\%, and power consumption increases by $3.668\%$. From configuration $1$ to configuration $9$, the L3 miss rate increases by $20.4\%$, and power consumption is reduced by $7.704\%$.

\section{Inference and Conclusions}
\label{sec:inf}
This paper has done a performance study of shared LLC  exploiting the heterogeneity for the Asymmetric Multi-Core architecture.  We started our investigation by understanding the metrics that significantly affect the L2 and LLCs by varying the parameters such as Cache size and associativity against different replacement strategies. We realized that replacement policy plays a significant impact on overall cache performance. From our first part of the study, we observe that an increase in cache size and degree of associativity improves the hit rate of a cache. But we see that the improvement in hit rate is not the same in all the cases. So we can say that cache hit rate not only depends on its parameters like cache size, associativity, block size, etc. But also depend upon the replacement policy that we are using and the different memory access patterns that the given workload follows.
\begin{table}[!h]
\footnotesize
\centering
\caption{Comparision results for all configurations}
\label{tab:comp}
\begin{tabular}{|p{2.8cm}|p{2.3cm}|p{2.9cm}|}
% {|c|c|c|}
 \hline
 Configuration & $L3$ miss rate & Power Consumption   \\ 
 \hline
  Configuration $1$ to $6$ & $11.3$\%  (Decrease) & $10.54\%$  (Decrease)\\
    Configuration $6$ to $9$ & $29.08\%$  (Increase) & $3.154\%$  (Increase)\\
     Configuration $1$ to $9$ & $17.9\%$  (Increase) & $7.761\%$  (Decrease)\\
\hline
\end{tabular}

\end{table}
In the second part of our work, we understand Cache Hierarchy and simulate the AMP Architecture. We inculcated the heterogeneity in our 16 core AMP by varying the individual core operating frequency and the method of execution, that is, whether the core is In-order or Out-of-order. It is crucial to mention beforehand that throughout our investigation of AMPs and their cache performance, the DVFS (Dynamic Voltage and Frequency Scaling) is kept enabled. We observe that the power variations in our results must not be considered on absolute terms; instead, they should be regarded as relative to each architecture.  We begin our simulations with all cores operating on the same frequency, with each core sharing both the L2 and LL caches also have Out-of-order execution. Then we gradually partition the caches such that they are private to each core and make some cores simpler, like lowering the frequency values and making them In-order to save on power. Our investigation showed that Configuration 6 is an optimal choice as it saves power compared to Configuration 1 and has the best LL miss rate. As we know that off-chip memory accesses usually consume more latency than on-chip memory accesses.  

We tested our architectures on multi-threaded PARSEC and SPLASH2 benchmarks to have a real-world understanding. From Section \ref{section:config169}, we have observed that on average, from configuration $1$ to configuration $9$, there is a $17.9\%$ increase in $L3$ miss rate and $7.761\%$ decrease in power consumption. We further observe from configuration $1$ to configuration $6$ that both the $L3$ miss rate and power consumption reduces by $11.14\%$ and $10.54\%$ respectively. In contrast, from configuration 6 to configuration 9, the L3 miss rate increases by $17.9\%$, and power consumption increases by $3.154\%$. So configuration $6$ is giving a better performance as well as power. The above-mentioned comparison results for all the configurations are listed down in Table. \ref{tab:comp}. 

An immediate next step to our work could be to increase the number of cores and bring more levels of heterogeneity in the system, such as flexible cache designs, varying the reorder buffer sizes, and propose many more optimizations.

\textbf{Funding:} This research received no specific grant from any funding agency in the public, commercial, or not-for-profit sectors.

\textbf{Informed Consent: } Not Applicable to this article.

\textbf{Conflict of Interest: }On behalf of all authors, the corresponding author
states that there is no conflict of interest.

\textbf{Data Availability Statement:} All data generated or analysed during the
study are included in this article.

\textbf{Author Contribution:\\} Murali Dadi: Conception and design of the study, Analysis and/or interpretation of data, Writing – original draft, Writing – review \& editing. \\Shubhang Pandey: Conception and design of study, Analysis and/or interpretation of data, Writing – original draft, Writing – review \& editing.\\ Aparna Behera: Conception and design of study, Analysis and/or interpretation of data, Writing – original draft, Writing – review \& editing.\\ TG Venkatesh: Conception and design of study, Analysis and/or interpretation of data, Writing – original draft, Writing – review \& editing.

%\section{Conclusion}
%\label{sec:con}
%AMPs have shown promise as the future architectures to solving several current computer architecture research problems. Investigations on proper task allocations to each core and even more diverse Cache Coherence Protocols could improve the performance of the existing systems. Also, the immediate next step to our work could be to increase the number of cores and bring more levels of heterogeneity in the system, such as flexible cache designs, varying the reorder buffer sizes, and propose many more optimizations.

% \bibliographystyle{IEEEtran}
\bibstyle{IEEEtran}
\bibliography{murali_bibtex}

\end{document}